\documentclass{PoS}
\usepackage{amsmath,amssymb,amsfonts,dsfont}
\usepackage{slashed}
\usepackage{cite}

\def\hbar{\hspace{0pt}\raisebox{1pt}{$-$} \hspace{-7pt} h}

\def\5{\overline 5}

\newcommand{\be}{\begin{equation}}
\newcommand{\ee}{\end{equation}}
\newcommand{\bea}{\begin{eqnarray}}
\newcommand{\eea}{\end{eqnarray}}

\newcommand{\ba}{\begin{eqnarray}}
\newcommand{\ea}{\end{eqnarray}}

\newcommand{\LL}{\mathrm{L}}

\newcommand{\TC}{\mathrm{TC}}
\newcommand{\SU}{\mathrm{SU}}
\newcommand{\Sp}{\mathrm{Sp}}
\newcommand{\U}{\mathrm{U}}

\newcommand{\EM}{\mathrm{em}}

\title{Revealing BSM composite dynamics
via topological interactions at future
colliders}

\ShortTitle{
Revealing BSM composite dynamics
via topological interactions at future
colliders
}

\author{\speaker{Natascia Vignaroli}\thanks{DNRF90, INFN}\\
        CP$^{3}$-Origins, University of Southern Denmark, Campusvej 55, DK-5230 Odense M, Denmark,\\
        INFN Sezione di Padova, via Marzolo 8, I-35131 Padova, Italy \\
        E-mail: \email{vignaroli@pd.infn.it}}

\author{Emiliano Molinaro \\
CP$^{3}$-Origins, University of Southern Denmark, Campusvej 55, DK-5230 Odense M, Denmark, \\
E-mail: \email{molinaro@cp3-origins.net}}
\author{Francesco Sannino \\
CP$^{3}$-Origins, University of Southern Denmark, Campusvej 55, DK-5230 Odense M, Denmark, \\
CERN Theoretical Physics Department, Geneva, Switzerland, \\
Danish IAS, University of Southern Denmark, Campusvej 55, DK-5230 Odense M, Denmark \\
E-mail: \email{sannino@cp3-origins.net} }
\author{Anders Eller Thomsen \\
CP$^{3}$-Origins, University of Southern Denmark, Campusvej 55, DK-5230 Odense M, Denmark \\
E-mail: \email{aethomsen@cp3.sdu.dk}}

\abstract{In composite Higgs models, new composite pseudoscalars can interact with the Higgs and with electroweak gauge bosons via anomalous interactions, which stem from the topological structure of the underlying theory. A future 100 TeV pp collider (FCC-pp) will be able to test these anomalous interactions and thus shed light on the strong dynamics which generates the Higgs and other composite resonances. We will discuss the topological interactions of a minimal composite Higgs model with fermionic ultraviolet completion, based on the coset SU(4)/Sp(4). We will indicate the strategy to test these interactions at the FCC-pp and the expected reach.}

\FullConference{The European Physical Society Conference on High Energy Physics\\
		5-12 July, 2017\\
		Venice}

\begin{document}

\section{Introduction}

An attracting and natural explanation to the electroweak (EW) symmetry breaking involves a strong gauge dynamics beyond the Standard Model (BSM), from which the Higgs emerges as a composite pseudo Nambu-Goldstone-Boson (pNGB) state \cite{CompositeHiggs}. 
The most minimal realization of a composite Higgs model which admits an underlying fundamental fermionic matter is based on the global flavor symmetry breaking pattern $\text{SU}(4) \to \text{Sp}(4) \sim\text{SO}(6) \to \text{SO}(5)$.
Besides the Higgs, the theory includes another composite pNGB, a CP-odd state $\eta$, and a pseudoscalar particle associated with the anomalous breaking of a global $\text{U}(1)$ symmetry in the new strong sector, analogous to the $\eta^{\prime}$ in QCD.
This class of theories is characterized by a topological structure which generates anomalous Wess-Zumino-Witten (WZW) \cite{Wess:1971yu
} interactions among the composite scalars and the gauge bosons. WZW terms depend on fundamental parameters of the composite theory, as the embedding of the EW sector in the coset and the number of degrees of freedom characterizing the new strong sector. Therefore, a study of the phenomenology associated with the topological interactions permits to uncover fundamental details of the underlying dynamics. This is similar to the case of the anomalous decay $\pi_0 \to \gamma\gamma$ which allowed for determining the number of colors in QCD. \\

We will present in this proceeding the main results of the study in \cite{Molinaro:2017mwb}, which analyzes the WZW interactions involving pseudoscalar composite states in the 
$\SU(4)/ \Sp(4)$ composite Higgs model and shows that they play a major role in their production at the planned future circular proton-proton collider with a beam-colliding energy of 100 TeV, the FCC-pp \cite{Arkani-Hamed:2015vfh}. The dominant processes are given by the vector-boson-fusion (VBF) production mechanism in association with a $W/Z$ boson or an Higgs in the final state. We will review the search strategy, delineated in \cite{Molinaro:2017mwb}, for detecting the process $\eta/\eta^{\prime}$ followed by a $W$ at the FCC-pp and present the collider reach on the fundamental parameters of the new strongly interacting theory.

\section{WZW terms in a SU(4)/Sp(4) composite Higgs model}

We consider a composite Higgs model with an underlying fundamental fermionic matter theory based on the coset $\text{SU}(4)/\text{Sp}(4)$.
The four fundamental technifermions, $\psi^a$, are in a pseudo-real representation, $R$, of a new strong gauge group $G_\TC$ and are assigned to vectorial representations of the SM gauge group $\SU(3)_\text{c}\times\SU(2)_\LL \times \U(1)_Y$, $(1,2)_0 \oplus (1,1)_{-1/2} \oplus (1,1)_{+1/2}$, in order to avoid gauge anomalies.
$\SU(4)$ is dynamically broken to $\Sp(4)$ via the fermionic bilinear condensate
\begin{equation}
	\left\langle \psi^{a} \epsilon_{\TC} \psi^{b} \right\rangle = f^2 \Lambda \,\Sigma_0^{ab} \, ,
	\label{eq:fermion_condensate}
\end{equation}
where $\Lambda\approx 4\pi f$ is the typical composite scale of the theory\footnote{Notice that this represents a simple estimate and should not be taken as a rigorous cutoff of the composite theory, as a precise evaluation requires non-perturbative methods.} and $f$ denotes the corresponding pion decay constant. The unitary matrix $\Sigma_0$ is in a two-index, antisymmetric representation of  $\SU(4)$, and takes the general form~\cite{Galloway:2010bp}
\begin{equation}
	\Sigma_0 = \cos\theta\, \Sigma_B \,+\, \sin\theta \,\Sigma_H = \begin{pmatrix}	i\,\sigma_2 \cos\theta  & \mathbf{1}\sin\theta \\ - \mathbf{1}\sin\theta & -i\,\sigma_2 \cos\theta \end{pmatrix}\,,\label{eq:vacuum}
	\end{equation}   
where $\theta$  parametrizes the alignment between the two physically inequivalent vacua of the theory, $\Sigma_B$ and
$\Sigma_H$ \cite{Cacciapaglia:2014uja}. 
The direction $\theta=0$ (the vacuum $\Sigma_B$) preserves the EW symmetry, while for $\theta=\pi/2$ (the vacuum $\Sigma_H$) the EW symmetry is fully broken (technicolor limit). 
The physical value of the vacuum alignment angle $\theta$ will depend on the effective potential of the
 theory below $\Lambda$, which is generated by the sources of explicit $\SU(4)$  breaking. 
 The EW scale is then given by
 \begin{equation}
v \;=\; 2 \,\sqrt{2}\, f\, \sin \theta\;=\;246~\text{GeV}\,.
\end{equation}
 As a general feature of composite Higgs models \cite{Contino:2010rs}, the natural minimum of the potential occurs either at $\theta=0$ (driven by gauge contributions) or at $\theta=\pi/2$ (from the top interactions with the new strong sector). In order to achieve the EW symmetry breaking and a light 125 GeV Higgs, it is necessary to operate a fine-tuned cancellation among different terms in the potential which misalign the vacuum to $0 < \sin\theta \ll 1$. Thus, $\sin\theta$ gives an estimate of the size of fine-tuning in composite Higgs models. 

Below the condensation scale $\Lambda$, the theory consists of composite states among which there are two pNGBs, the Higgs and the pseudoscalar $\eta$ (the three remaining pNGBs become the longitudinal polarizations of the $W/Z$ bosons after the EW symmetry breaking).   
The composite scalars undergo anomalous interactions with the EW gauge bosons, which are described by the renowned gauged Wess-Zumino-Witten (WZW) action \cite{Wess:1971yu
} and depend on fundamental parameters of the underlying gauge dynamics: $\sin\theta$ and the dimension of the technifermion representation under $G_\TC$, $d(R)$.
The WZW action leads to the following $\eta$ interactions with two gauge bosons (similar terms can be found for the $\eta ^ \prime$ composite pseudoscalar \cite{Molinaro:2017mwb}):
\begin{equation}
	 -\dfrac{d(R) \alpha_\EM \cos\theta \sin\theta}{32\pi v} \eta \left[\dfrac{2}{c_w s_w} A_{\mu\nu}\tilde{Z}^{\mu\nu} +\dfrac{c_w^2 -s_w^2 }{c_w^2 s_w^2} Z_{\mu\nu} \tilde{Z}^{\mu\nu} +\dfrac{2}{s_w^2} W^+_{\mu\nu} \tilde{W}^{-\mu\nu} \right] \ .
	\label{eq:WZW-AAeta}
\end{equation}
By expanding the WZW action up to dimension-6 operators (note that this has been realized for the first time in \cite{Molinaro:2017mwb}), terms where composite pseudoscalars interact with three gauge bosons or with two gauge bosons and the Higgs are generated:
\begin{equation}
	\begin{split}
	& -i\dfrac{d(R) \alpha^{3/2}_\EM\cos\theta \sin\theta}{4\sqrt{\pi} v} \varepsilon^{\mu\nu\rho\sigma} \partial_\mu\eta \left[\dfrac{2}{s_w^2}A_\nu + \dfrac{2c_w^2 - \sin^2\theta}{c_w s_w^3}Z_\nu \right] W_\rho^+ W_\sigma^- ,
	\end{split}
	\label{eq:WZW-AAAeta}
\end{equation}

\begin{equation}\label{eq:h}
	\begin{split}
	&- \dfrac{d(R)\alpha_\EM\sin^3 \theta }{16 \pi v^2} \varepsilon^{\mu\nu\rho\sigma} \bigg[\dfrac{4}{c_w s_w} \partial_\mu h \partial_\nu \eta A_\rho Z_\sigma \\
	&+ h\overleftrightarrow{\partial_\mu} \eta
	\left( \dfrac{c_w^2-s_w^2}{c_w^2 s_w^2}Z_{\nu\rho}Z_\sigma +\dfrac{1}{s_w^2} \left(W_{\nu\rho}^+ W_\sigma^- + W_{\nu\rho}^- W_\sigma^+ \right) +\dfrac{1}{c_ws_w} \left(A_{\nu\rho}Z_\sigma + Z_{\nu\rho}A_\sigma \right) \right)\bigg].
	\end{split}
	\end{equation}
Notice that, being the theory free of gauge anomalies, anomalous triple and quartic gauge couplings are not generated.

\section{
Phenomenology of composite pseudoscalars at the LHC and at the FCC-pp}

\begin{figure}[t!]
\begin{center}
\includegraphics[width=0.47\textwidth]{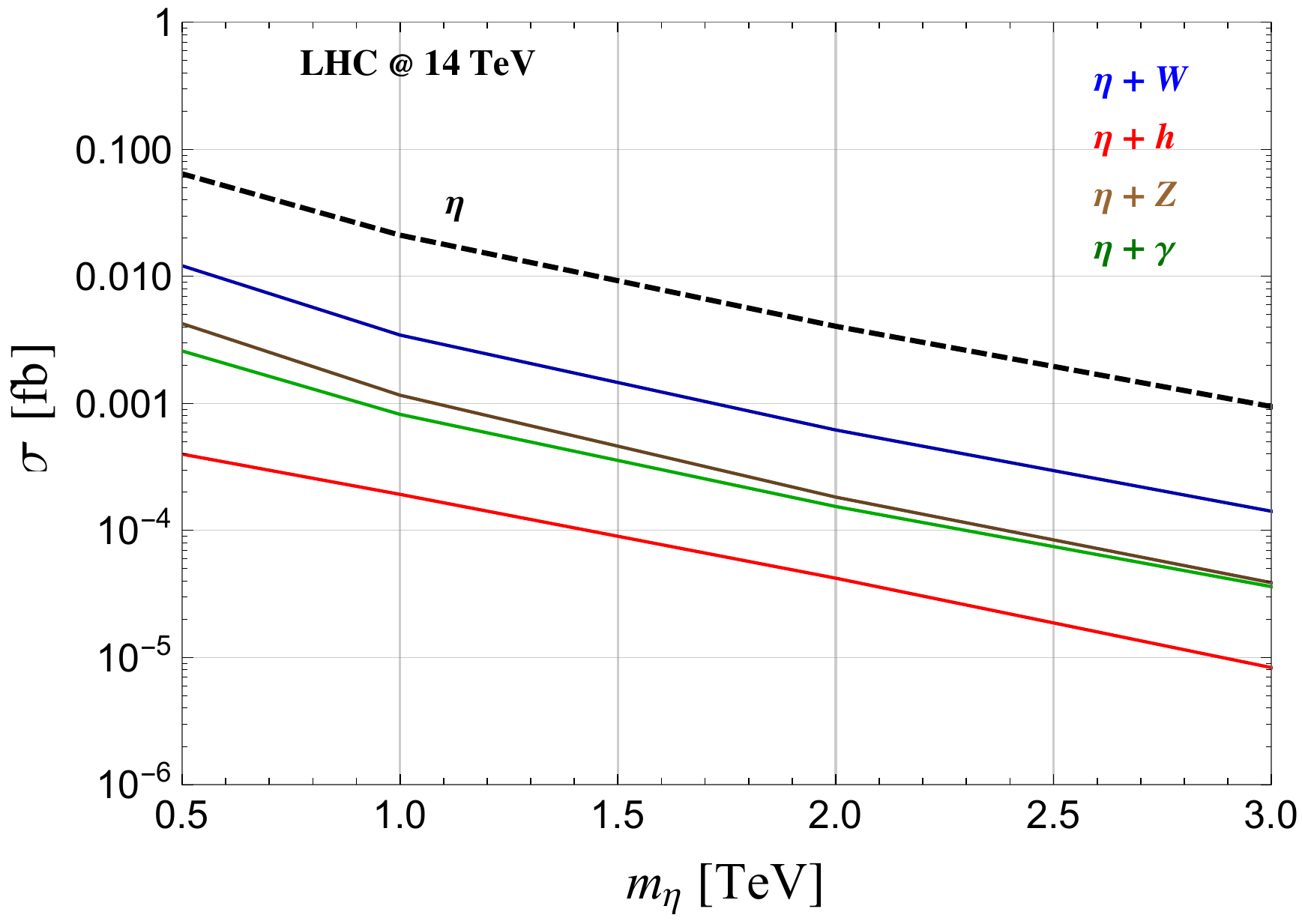}\quad \includegraphics[width=0.47\textwidth]{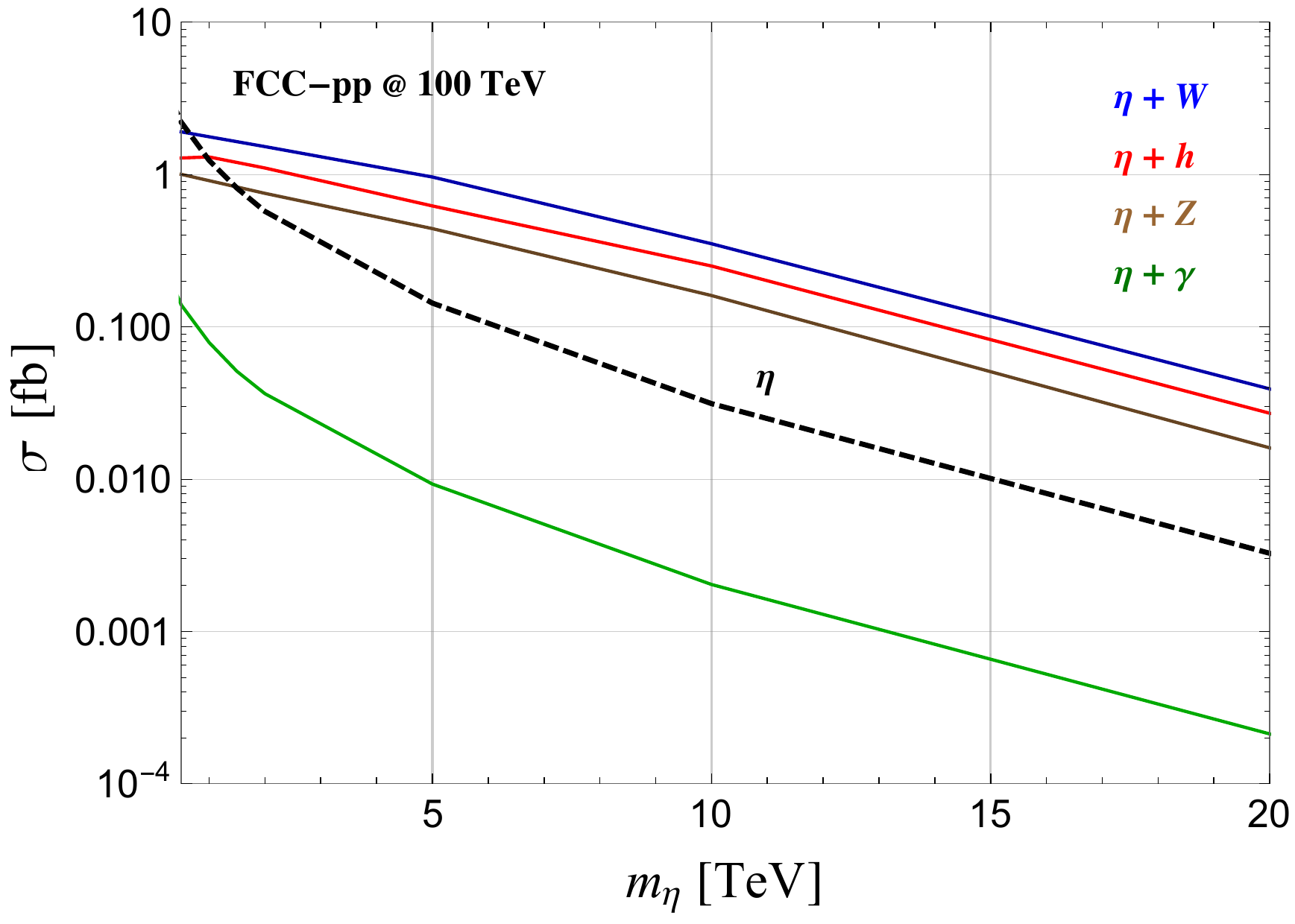} \end{center}
\caption{ \label{fig:xsec}
Cross sections for different $\eta$ production mechanisms at the 14 TeV LHC ({\it left panel}) and at the 100 TeV FCC-pp ({\it right panel}). The black-dashed curve indicates the VBF production resulting only from the pseudoscalar topological interaction with two gauge bosons (eq.~(\ref{eq:WZW-AAeta})). The cross section for the $\eta$ production accompanied by the Higgs or by a gauge boson results from the interference between processes mediated by topological interactions of the pseudoscalar with two (eq.~(\ref{eq:WZW-AAeta})) and three particles (eqs.~(\ref{eq:WZW-AAAeta}) and (\ref{eq:h})). We fix $d(R)=10$ and $\sin\theta =$ 0.25.  }
\end{figure}

\begin{figure}
\begin{center}
\includegraphics[width=0.3\textwidth]{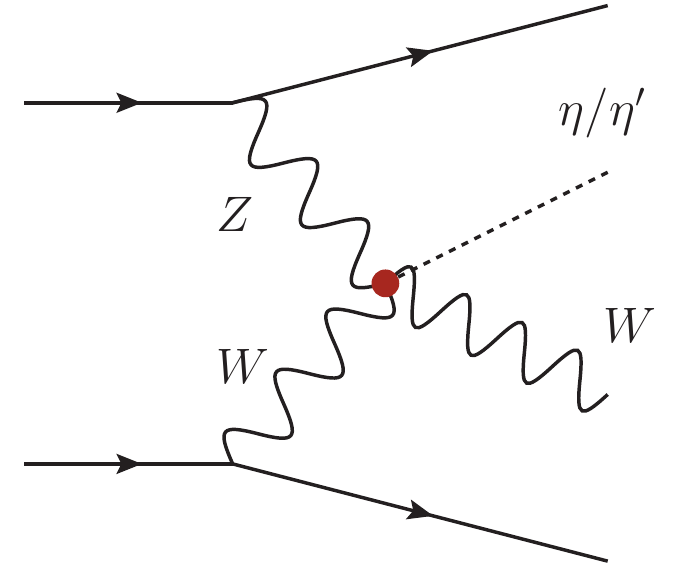}
 \hspace{1cm}
\includegraphics[width=0.3\textwidth]{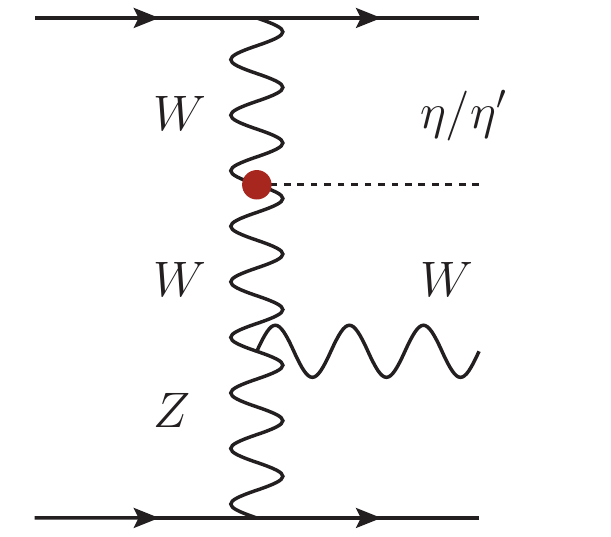}
\caption{ \small Relevant diagrams for the production via VBF of an $\eta/\eta^\prime$ particle accompanied by a $W$ boson.}
\label{fig:S-diagram}
\end{center}
\end{figure}

If the $\eta$ (similar considerations apply to the $\eta^\prime$) does not couple directly to fermions, the dominant production mechanism is vector-boson-fusion (VBF) mediated by topological interactions. The $\eta$ can also be doubly-produced via VBF through non-topological kinetic interactions. The cross sections for these processes at the LHC are small.  Figure \ref{fig:xsec} shows the cross sections for different $\eta$ production mechanisms at the 14 TeV LHC (left panel) and at the 100 TeV FCC-pp (right panel), as a function of the pseudoscalar mass. 
The black-dashed curve shows the VBF production of the $\eta$ resulting only from the pseudoscalar topological interactions with two gauge bosons (eq.~(\ref{eq:WZW-AAeta})). The colored continuous curves indicate the $\eta$ production accompanied by the Higgs or by a gauge boson. These processes are mediated by topological interactions with four bosons (eqs.~(\ref{eq:WZW-AAAeta}) and (\ref{eq:h})), which interfere with the terms in the WZW action involving only three bosons (eq. (\ref{eq:WZW-AAeta})).
Our calculations show that the  VBF associated production with an EW gauge boson or a Higgs represent the dominant production mechanisms of the pseudoscalars at the 100 TeV collider. These processes are highly enhanced at the FCC-pp compared to the LHC, as an effect of the derivative coupling structure of the vertices entering in diagrams similar to those in Fig. \ref{fig:S-diagram}. 

\section{FCC-pp reach on topological interactions}

\begin{figure}[t!]
\begin{center}
\includegraphics[width=0.48\textwidth, height=5.3cm]{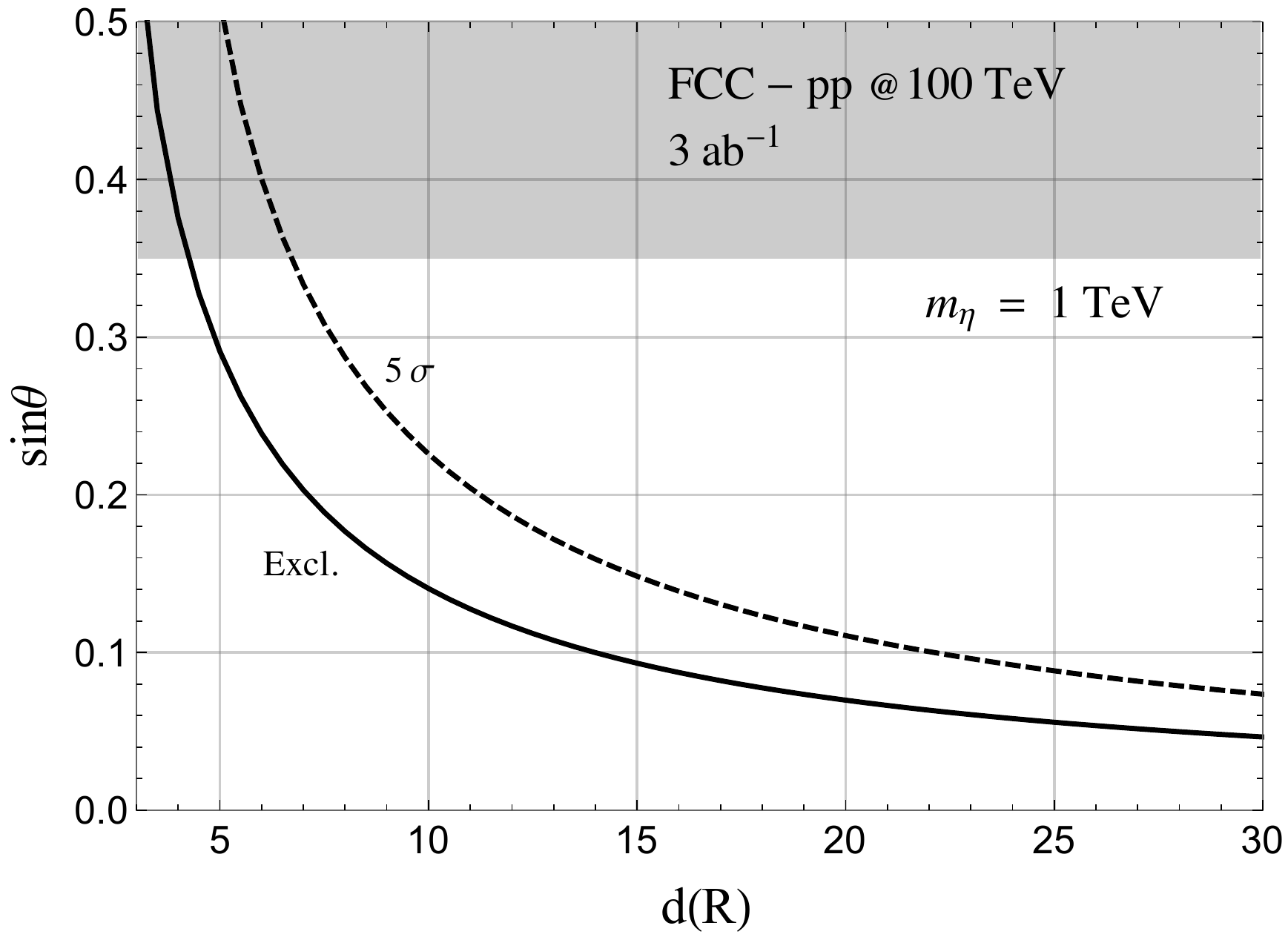}\quad
\includegraphics[width=0.48\textwidth, height=5.3cm]{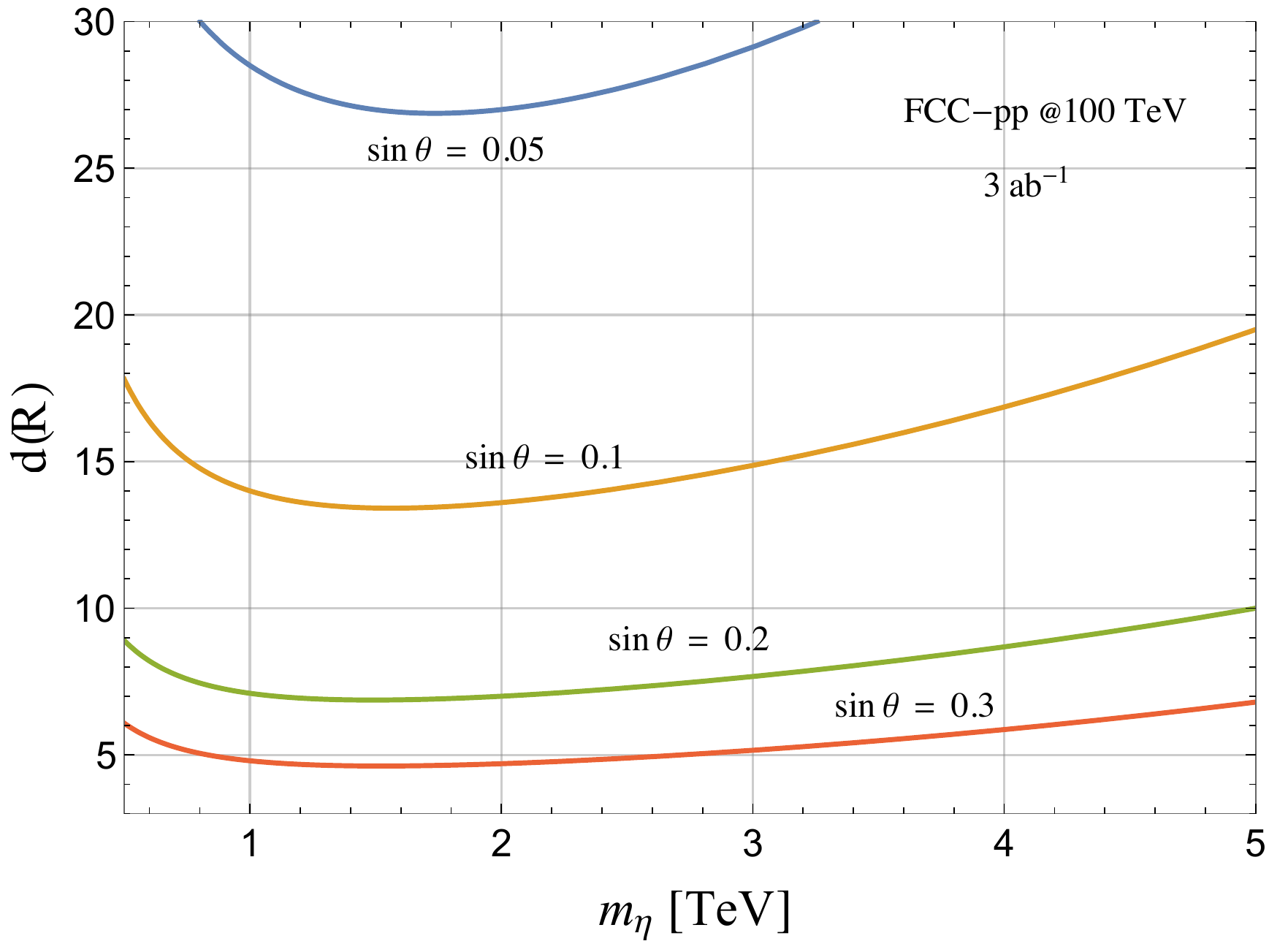} 
\caption{ \small Reach at a 100 TeV proton-proton collider with a luminosity of 3 ab$^{-1}$. {\it Left panel}: 5$\sigma$ (dashed curve) discovery and 95\% CL exclusion (continuous line) reach for the signal with an $\eta$ of 1 TeV in the plane ($d(R)$, $\sin\theta$). The shaded area is excluded by LHC Higgs data if the Higgs Yukawa coupling to the top is modified as $\cos\theta y^{SM}_t$. {\it Right panel}: 95\% CL exclusion curves for different $\sin\theta$ values on the plane ($m_{\eta}$, $d(R)$). }
\label{fig:reach-eta}
\end{center}
\end{figure}

We performed in \cite{Molinaro:2017mwb} a detailed analysis of the $\eta/\eta^\prime$ production followed by a $W$ at the 100 TeV FCC-pp (Fig.  \ref{fig:S-diagram}).  
We considered the case where the $\eta$ does not directly couple to SM fermions while the $\eta^{\prime}$ interacts with the top quark, as resulting from the scenario where the top mass is generated from a 4-fermion operator. The $\eta^{\prime}$ thus decays dominantly to $t\bar{t}$, while the $\eta$ decays to $W^{+}W^{-}$, $ZZ$ and $Z\gamma$ through the topological interactions in (\ref{eq:WZW-AAeta}). \footnote{In the case of very massive pseudoscalars, $m_{\eta/\eta^{\prime}}\gtrsim 5$ TeV, the 3-body decays into gauge bosons, mediated by 3- and 4-boson topological interactions, become relevant (see, e.g., \cite{Molinaro:2016oix 
}).} 
We considered the subsequent fully-hadronic decay of the pseudoscalars and applied a jet reconstruction procedure, based on an anti-$k_T$ algorithm with jet cone size $R=1.5$, which identifies a (mostly) single fat-jet coming from the pseudoscalars in the case where they directly decay either to gauge bosons (via topological terms) or to tops. 
The final state we considered is thus
\begin{equation} \label{eq:final-state}
\ell+ n \, \text{jets} \, + \slashed{E}_T \, \, , \qquad n \geq 3 \, ,\qquad \ell \equiv e,\mu \ ,
\end{equation}
where $\slashed{E}_T$ indicates the transverse missing energy. 
The signal is characterized by a peculiar kinematics, typical of a VBF process, with two final forward-backward jets emitted at high rapidity \cite{Golling:2016gvc, Mohan:2015doa, Goncalves:2017gzy}. The produced $W$ and pseudoscalar have large transverse momenta and their decays produce large missing energy, a large-$p_T$ lepton and a hard $p_T$-leading fat-jet. By exploiting these features it is possible to disentangle the signal from the background, which is dominantly made of $W$+jets events.  The FCC-pp reach is then evaluated after a cut and count analysis on simulated signal and background samples. The simulations are performed with MG5\_aMC@NLO \cite{Alwall:2014hca} plus PYTHIA 6.4 \cite{Sjostrand:2006za}, for showering and hadronization, and including a smearing to the jet energy, 
  in order to mimic detector effects.
We refer the reader to \cite{Molinaro:2017mwb} for details on the cuts applied and on the $\eta/\eta^\prime$ reconstruction procedure.

\section{Conclusions}

The analysis of the $\eta/\eta^\prime$ production followed by a $W$ allows us to indicate the FCC-pp reach on the main parameters characterizing the underlying strong dynamics, $d(R)$ and $\sin\theta$.
Our final results for the signal with the $\eta$ (we obtain a slightly reduced reach for the signal with the $\eta^\prime$) are presented in Fig. \ref{fig:reach-eta}.
We see that the FCC-pp can test the underlying gauge dynamics for $d(R)$ as small as 13 (5) and $\sin\theta=0.1 \, (0.3)$.
Our analysis is also a promising discovery search for the $\eta/\eta^\prime$ resonances at the FCC-pp, in the case they were not already discovered at the LHC, due to the low production cross section.
We conclude that our findings in \cite{Molinaro:2017mwb} show that new channels with topological interactions, in particular the pseudoscalar production via VBF in association with a $W/Z$ or a Higgs boson, are very promising to test strong dynamics beyond the SM related to the mechanism of EW symmetry breaking.

\end{document}